# A Universal Model for Bingham Fluids with Two Characteristic Yield Stresses

N.G.Domostroeva[1] and N.N.Trunov[2]

*D.I.Mendeleyev Institute for Metrology*

*Russia, St.Petersburg. 190005 Moskovsky pr. 19*

February, 4, 2009

**Abstract:** We introduce a new model for Bingham fluids with a complicated dependence of the deformation velocity on the shear stress. Specific features of the flow rate for such fluid are studied. This model embraces many special and limiting cases, for many of them analytic calculations become possible.

## 1. Approximate universal model

As it turned out recently, many important fluids including oil reveal a more complicated rheological behavior than it is usually postulated for a Bingham fluid [1]. For a uni-directional shear flow with the velocity

$$v_x = v(y) \tag{1}$$

the following dependence of the deformation velocity

$$\dot{\gamma} = \frac{dv}{dy} \tag{2}$$

on the shear stress $\tau$ takes place, see figure 1.

---

[1] Electronic address: N.G.Domostroeva@vniim.ru
[2] Electronic address: trunov@vniim.ru

Namely,

$$\dot{\gamma} = 0, \text{ if } \tau < \tau_0 \qquad (3)$$

and asymptotically

$$\dot{\gamma} = \frac{\tau - \tau_d}{\mu} \qquad (4)$$

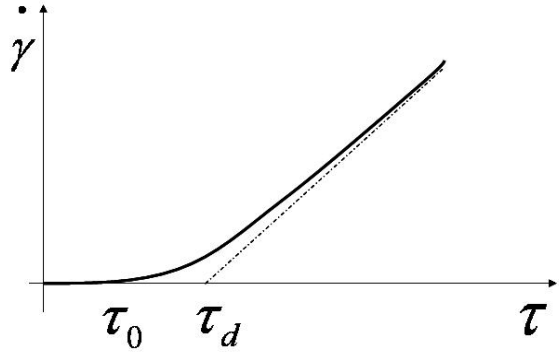

Fig.1

for large values $(\tau - \tau_d)$ exceeding roughly (1-2) $\tau_d$. Thus we have two characteristic values of the shear stresses: the yield stress $\tau_0$ and a parameter $\tau_d > \tau_0$ related to the asymptotic viscosity $\mu$. In the intermediate interval between (2) and (3) there is a flow with a very high effective viscosity.

If $\tau - \tau_0 \ll \tau_0$, we have

$$\dot{\gamma} = \frac{\tau - \tau_0}{\mu_0} \qquad (5)$$

with a very large $\mu_0 \gg \mu$.

For the quantative description of such fluid we have to find a simple enough universal and physically grounded model function $\gamma(\tau)$. This function depends on four parameters $(\tau_0, \mu_0, \tau_d, \mu)$ or $(\tau_d, \mu, \alpha, \beta)$

$$\alpha = \frac{\mu}{\mu_0} < 1, \quad \beta = \frac{\tau_0}{\tau_d} < 1 \qquad (6)$$

with dimensionless $\alpha, \beta$.

As it can be seen from fig.1, this function must obey two conditions

$$\frac{d\dot\gamma}{d\tau} > 0, \quad \frac{\partial^2 \dot\gamma}{\partial \tau^2} > 0 \qquad (7)$$

and have asymptotics (3) and (4). Besides, for a usual Newtonian fluid with $\tau_0 = \tau_d = 0$ such a function must be odd:

$$\gamma(\tau) = -\gamma(-\tau) \qquad (8)$$

A convenient function satisfying all the above conditions is

$$\gamma(\tau) = \frac{1}{\mu}\left[\tau - \tau_0 - (\tau_d - \tau_0)\tanh\frac{(1-\alpha)(\tau - \tau_0)}{\tau_d - \tau_0}\right] \qquad (9)$$

for all $\tau \geq \tau_0$ with $\alpha$ (6). In particular, for small $\tau - \tau_0$ replacing tanh(z) by its argument z we obtain:

$$\gamma = \frac{(\tau - \tau_0)\alpha}{\mu} = \frac{(\tau - \tau_0)}{\mu_0} \qquad (10)$$

## 2. Flow in a channel

Now we study the flow with $\gamma(\tau)$ (9) in a channel with infinite width and the height $2h$, so that

$$-h \leq y \leq h \qquad (11)$$

Obviously

$$v(h) = v(-h) = 0 \qquad (12)$$

In the upper half

$$\tau = gy, \quad g = dp/dx \qquad (13)$$

with $g$ being the pressure gradient. Substituting (13) into (9) and integrating we obtain

$$v = \frac{(h-y_0)^2 - (y-y_0)^2}{2} + \frac{z^2}{1-\alpha} \ln \frac{\cosh(1-\alpha)(y-y_0)/z}{\cosh(1-\alpha)(h-y_0)/z} \tag{14}$$

if $y \geq y_0$ and

$$v(y) = v(y_0), \text{ if } y \leq y_0 \tag{15}$$

Here we have introduced

$$y_0 = \frac{\tau_0}{g}, \ y_d = \frac{\tau_d}{g}, z = y_d - y_0 \tag{16}$$

The total flow rate ( for both halves per unit width) is

$$Q = 2\int_{y_0}^{h} v(y)dy + 2y_0 v(y_0) \tag{17}$$

Hereafter we use the equality

$$\ln \frac{\cosh A}{\cosh B} = A - B + \ln \frac{1+\exp(-2A)}{1+\exp(-2B)} \tag{18}$$

We suppose that $2A$, $2B$ in (18) exceed 2-3. In this case we may assume that for the logarithmic terms the upper limit of integration in (17) is equal to infinity and use

$$\int_0^\infty ds \ln(1+e^{-2sk}) = \frac{\pi^2}{24k} \tag{19}$$

The final expression looks as follows :

$$Q = Q_0 \left[1 - \frac{3}{2}(a+b) + \frac{1}{2}\left(a^3 + 3a^2b + \frac{\pi^2 b^3}{4(1-\alpha)^2}\right)\right] \tag{20}$$

with

$$Q_0 = \frac{2h^3 g}{3\mu}; \quad a = \frac{\tau_0}{hg}; \quad b = \frac{\tau_d - \tau_0}{hg} \tag{21}$$

The well known standard expression for $Q$ we obtain if $b = 0$ i.e. $\tau_0 = \tau_d$. Then for all $\tau \geq \tau_0 = \tau_d$ the viscosity is equal to $\mu$ so that $\mu_0$ does not have influence on $Q$.

Introducing a new parameter:

$$c \equiv \frac{\tau_d}{hg} = a + b \tag{22}$$

so that

$$a = \beta c, \quad 0 \leq \beta \leq 1, \tag{23}$$

we obtain another convenient formula:

$$\frac{Q}{Q_0} = 1 - \frac{3c}{2} + \frac{c^3}{2} f(\alpha, \beta), \tag{24}$$

$$f(\alpha, \beta) = \beta^3 + 3\beta^2(1-\beta) + \frac{\pi^2(1-\beta)^3}{4(1-\alpha)^2} \tag{25}$$

The abovementioned standard case corresponds to $\beta = 1$, when $f = 1$ and $Q = 0$ if $c = a \leq 1$. Remember that in many typical cases $a^3$ is much smaller than $a$ so that the term with $a^3$ may be omitted.

The following features of Eq. (24) are remarkable:

1. Both the first term and the second, linear in $c$ term do not depend on $\beta = \tau_0/\tau_d$;
2. there is no term of order $c^2$ at any $\beta$;
3. only the smallest and often neglected term proportional to $c^3$ depends on $\alpha$.

The same calculations were fulfilled with several another model functions $\dot{\gamma}$ satisfying to the previous conditions. In all these cases the above properties 1)-3) remain valid and only the specific form of $f$ depends on $\dot{\gamma}(\tau)$.

The usual way for determining the yield stress takes into account only the linear in $c$ asymptotics of $Q$ (25). It is evidently that in such a way we can only determine $\tau_d$ and not $\tau_0$.

Thus namely $\tau_d$ and not $\tau_0$ remains a genuine parameter determining the whole rheology. As to $\tau_0$, it is in the presence of $\tau_d$ a "dumb" parameter for all or the most of situations. Thus we cannot agree with the opinion [1] that $\tau_d$ is only a formal parameter.

### 3. Possible applications

Thus we have a rather simple bi-viscosity model (9) including limiting cases: $\mu_0 = \infty$, if $\alpha = 0$ and so on. Moreover, a very weak dependence of the whole rheology on $\tau_0$ in the presence of $\tau_d$ allows us in most cases simply to put $\tau_0 = 0$. Then we obtain a fully analytic approximate form (9) which is more physically and mathematically well grounded than the well known power law approximation: $\dot{\gamma}$ proportional to $\tau^m$ [1,4] for the Bingham fluid. Our form of $\dot{\gamma}$ may be used for analytical calculation of the flow in complex geometry [2] e.g. in a squeeze test [3].

1. V.V.Tetelmin, V.A Yasev. *The Oil Reology, Moscow,* 2009 [in Russian], p.43.
2. G.G.Lipcomb, M.M.Denn. *J.Non-Newton Fluid Mech.*, 14 (1984) PP 337-346.
3. N.Roussel, Ch.Lanos, Z.Toutou. *J.Non-Newton Fluid Mech.*,135 (2006) 1-7.
4. G.Astarita, G.Marrucci. *Principles of Non-Newtonian Fluid Mechanics. McGraw-Hill,* 1974, ch. 2.